\documentclass[aip,apl,reprint,a4paper,amssymb,amsmath,amsfonts]{revtex4-1}

\usepackage{graphicx}
\usepackage{hyperref}

\begin{document}

\title{Strain control of electron-phonon energy loss rate in many-valley semiconductors}

\author{J. T. Muhonen}
\email{juha.muhonen@tkk.fi}
\affiliation{Department of Physics, University of Warwick, CV4 7AL, UK}
\affiliation{Low Temperature Laboratory, Aalto University, P.O. Box 13500, FI-00076 AALTO, Finland} 

\author{M. J. Prest}
\affiliation{Department of Physics, University of Warwick, CV4 7AL, UK}

\author{M. Prunnila}
\affiliation{VTT Technical Research Centre of Finland, P.O. Box 1000, FI-02044 VTT, Espoo, Finland}

\author{D. Gunnarsson}
\affiliation{VTT Technical Research Centre of Finland, P.O. Box 1000, FI-02044 VTT, Espoo, Finland}

\author{V. A. Shah}
\affiliation{Department of Physics, University of Warwick, CV4 7AL, UK}

\author{A. Dobbie}
\affiliation{Department of Physics, University of Warwick, CV4 7AL, UK}

\author{M. Myronov}
\affiliation{Department of Physics, University of Warwick, CV4 7AL, UK}

\author{R. J. H. Morris}
\affiliation{Department of Physics, University of Warwick, CV4 7AL, UK}

\author{T. E. Whall}
\affiliation{Department of Physics, University of Warwick, CV4 7AL, UK}

\author{E. H. C. Parker}
\affiliation{Department of Physics, University of Warwick, CV4 7AL, UK}

\author{D. R. Leadley}
\affiliation{Department of Physics, University of Warwick, CV4 7AL, UK}

\begin{abstract}
We demonstrate significant modification of the electron-phonon energy loss
rate in a many-valley semiconductor system due to lattice mismatch induced
strain. We show that the thermal conductance from the electron system to the
phonon bath in strained n$^{+}$Si, at phonon temperatures between 200 mK and
450 mK, is more than an order of magnitude lower than that for a similar
unstrained sample.
\end{abstract}

\date{\today}
\maketitle

The interaction between electrons and phonons is one of the most important parameters of a semiconductor system, because it dictates the room temperature electron mobility and electron-phonon (\textit{e-ph}) energy loss.\cite{ridley:1991rev} In the low temperature (sub-1 K) regime \textit{e-ph} coupling typically dominates the total thermal coupling from electrons to the environment. Therefore, \textit{e-ph} coupling is one of the key parameters of interest in low temperature bolometer and microcooler applications.\cite{giazotto:2006rmp} In single-valley semiconductors the electron-phonon interaction can be typically described by deformation potential or piezoelectric coupling constants, which means that the dependence of the \textit{e-ph} matrix elements on the electronic variables, such as momentum, can be typically ignored. \cite{ganthmakher:b1} The situation is quite different in metals where momentum dependency must be included explicitly.

\begin{figure}
\begin{center}
\includegraphics{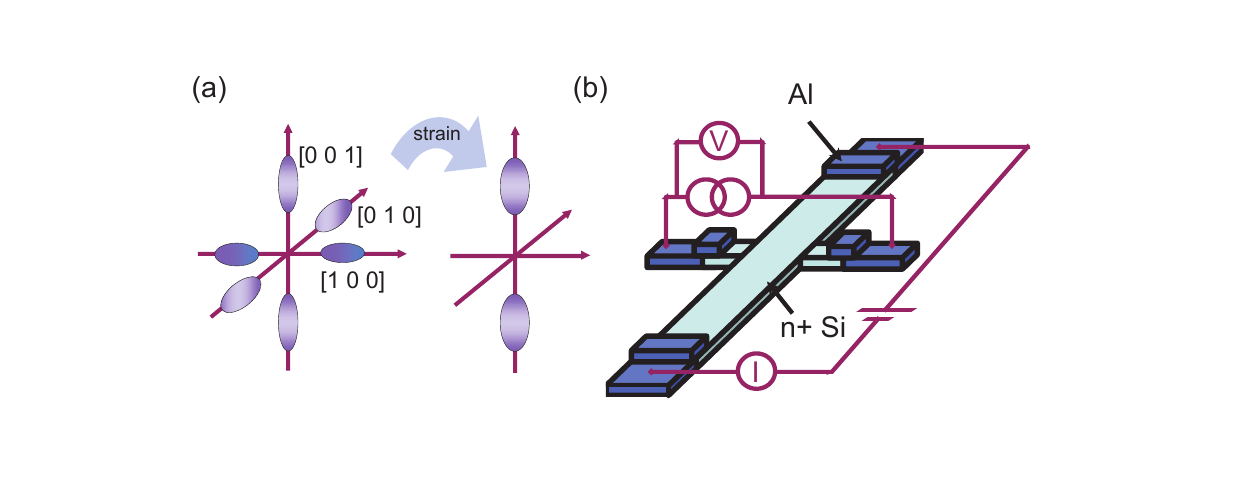}
\end{center}
\caption{(Color online) \textbf{(a)} Schematic illustration of the constant energy ellipsoid of Si conduction band valleys and how the strain should affect these. The four in-plane valleys have their energy raised and are depopulated due to strain.
\textbf{(b)} Simplified geometry of the sample and a sketch of the measurement setup, Al in dark and Si in light color. The center n$^{+}$ Si mesa is 205 $\protect\mu $m long and 5 $\protect\mu $m wide. We voltage bias the wire from the ends and measure the induced heating current $I$. Simultaneously the thermometer junctions are biased with a constant current and the voltage $V$ is monitored. This voltage gives the electron temperature $T_{e}$.}
\label{fig1}
\end{figure}

Simply setting the \textit{e-ph} matrix elements to a constant is not the whole story even for deformation potential coupling in semiconductors. Certain lattice perturbations can lift the valley degeneracy so that the valley degree of freedom plays an important role in low temperature \textit{e-ph} energy loss.\cite{prunnila:2005,prunnila:2007} Mean-field theory predicts that the \textit{e-ph} energy loss rate is strongly enhanced in many-valley semiconductors in comparison to single-valley ones due to lack of screening.\cite{prunnila:2007} This suggests that if one is able to depopulate asymmetric valleys of a many-valley semiconductor then a strong reduction in \textit{e-ph} energy loss rate should be observed. In this letter, we will experimentally test this prediction by using strain to depopulate the four in-plane valleys of epitaxial n$^{+}$Si films [see Fig.~\ref{fig1}(a)]. Sub-1 K \textit{e-ph} energy loss experiments of strained samples are compared to experimental data of Ref.~\onlinecite{prunnila:2005} and our non-strained control sample. We find that qualitatively the experiments agree with theory and \textit{e-ph} thermal conductance is indeed reduced in the strained sample. The effect is, however, smaller than ideally expected.

Strained silicon epi-layers were produced using a RP-CVD epitaxial growth system. Initially, a linearly graded Si$_{1-x}$Ge$_x$ alloy layer of thickness 2 $\mu$m is grown, with the germanium content (x) graded up to 0.2. This is followed by a constant composition layer of 500 nm Si$_{0.8}$Ge$_{0.2}$. Finally, an epitaxial silicon layer of 30 nm is grown, which conforms to the lattice spacing of the alloy layer and is thus tensile strained. This silicon layer has a phosphorus dopant concentration of $4 \times 10^{19}$ cm$^{-3}$. As a control, a similar doped silicon layer was grown on silicon without a SiGe virtual substrate. The degree of relaxation of the SiGe buffer layer and strain in the surface silicon layer were determined by X-ray diffraction. We found that the strain in the silicon layer is approximately 0.95\% and hence the band splitting between the 2-fold and 4-fold conduction bands should be about 150 meV \cite{Vogelsang:1993}, which is sufficient that, at the targeted carrier concentration, only the 2-fold bands are populated.

The geometry of the devices fabricated on these wafers is presented in Fig.~\ref{fig1}(b). Rectangular mesa pillars of height 100 nm (active layer 30 nm), were etched using a CF$_{4}$:O$_{2}$ (10:1) plasma to define the geometry of the degenerately doped material and isolate the mesas from the surrounding area. After native oxide removal in HF, the aluminum leads were then sputter deposited to a thickness of 200 nm and patterned using a phosphoric acid based etch solution.

The degenerately doped silicon region is connected to the aluminium (superconductor) leads through Schottky barriers. \cite{savin:2001} These kind of superconductor-semiconductor junctions exhibit similar characteristics as the well known superconductor-insulator-normal metal junctions.\cite{giazotto:2006rmp} The ratio of the subgap resistance to the asymptotic resistance, a common figure of merit for these kinds of junctions, is 200 for the strained sample and 40 for the control. The temperature dependency of the I-V characteristics of these junctions is used for thermometry in our experiment. We bias the two junctions in the middle of the bar with a constant current from a floating high impedance current source and monitor the induced voltage, which depends on the electronic temperature of the semiconductor island.\cite{savin:2001} The bias current through the thermometer junctions is selected so that it has a negligible heating effect. We calibrate these thermometers against the RuO thermometer of our dilution cryostat. These junction thermometers, however, saturate at around 200 mK, because of the relatively low subgap resistances. Hence, all the data we show are above this temperature. Sample parameters are presented in Table~\ref{tab1}.

\begin{table}
\caption{Sample parameters. Junction resistance and Si sheet resistance measured at 50 mK, carrier densities and mobilities from Hall measurements at 10 K.}
\centering
\begin{tabular}{|c|c|c|}
\hline
Sample & Strained & Control \\ \hline
Si-Al junction resistance (k$\Omega\mu$m$^2$) & 108 & 12 \\ \hline
Si sheet resistance ($\Omega$) & 569 & 383 \\ \hline
Carrier density ($10^{19}$ cm$^{-3}$) & 2.7 & 3.1 \\ \hline
Mobility (cm$^2$/V) & 155 & 192 \\ \hline
Mean free path (nm) & 5.3 & 5.0 \\ \hline
\end{tabular}
\label{tab1}
\end{table}

As shown in Fig.~\ref{fig1}(b), we induce heating current $I$ through the semiconductor island with voltage bias. The total heating power $P$ is calculated from $P=R_{\text{Si}}I^{2}$, where $R_{\text{Si}}$ is the resistance of the n$^{+}$Si mesa (measured in a 4-probe setup). We neglect the heating effect from the Al-Si junctions at the ends, assuming that because of the high resistivity of the n$^{+}$Si, and the distance of the thermometer from the junctions, the total electron-phonon coupling is much stronger than the heat conductivity through the wire. Hence, the junction heating is effectively \textquotedblright screened\textquotedblright by the electron-phonon coupling. We checked this assumption with finite-element-method simulations taking electronic thermal conductivity from the Wiedemann-Franz law and using our measured electron-phonon couplings (see below). The effect of junction heating on the electronic temperature near the electron thermometer was less than 1 $\%$ in all cases.

\begin{figure}
\begin{center}
\includegraphics{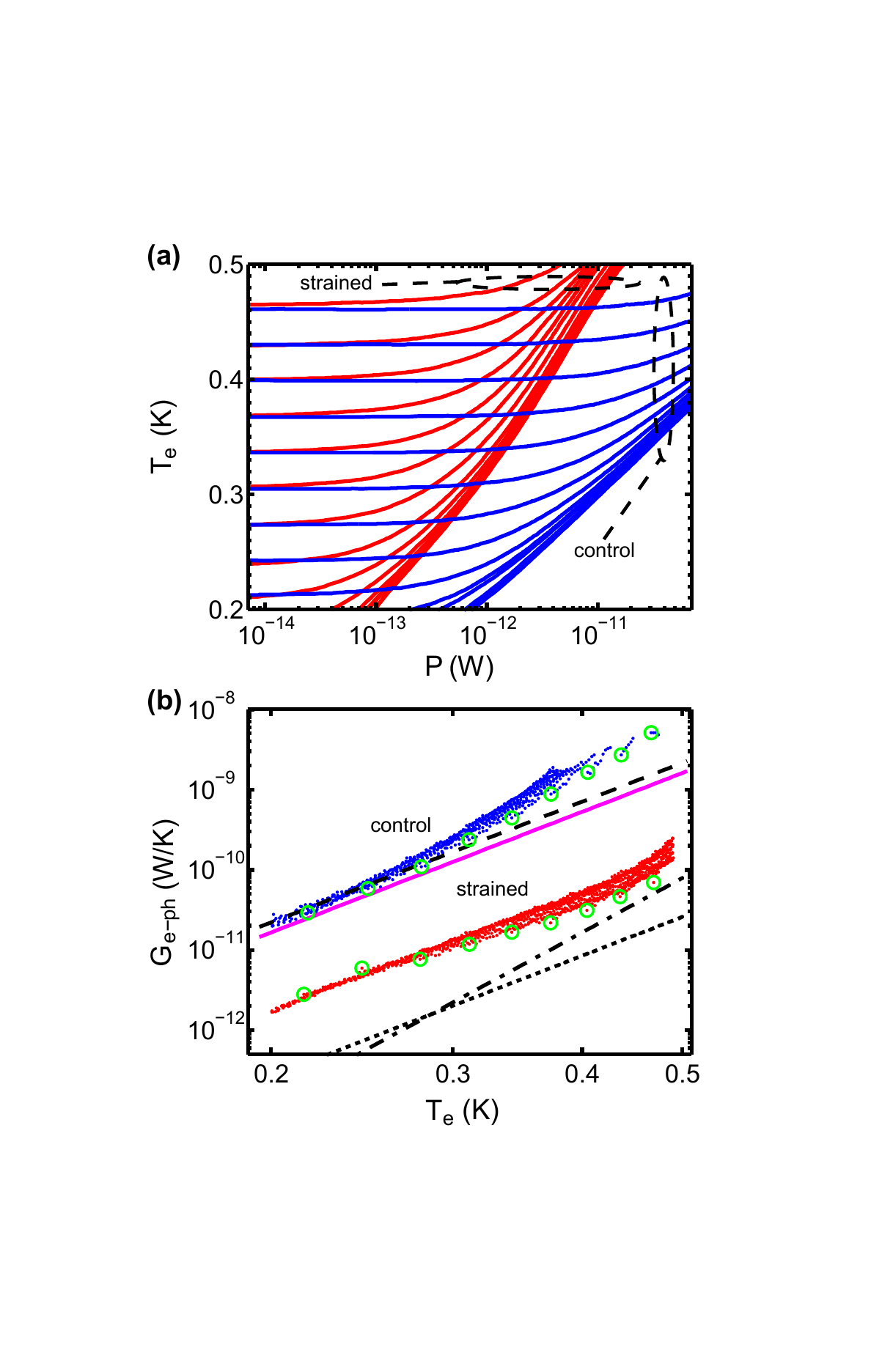}
\end{center}
\caption{(Color online) \textbf{(a)} Measured electronic temperature of the control and strained sample as a function of the heating power applied at different bath temperatures.
\textbf{(b)} The electron-phonon thermal conductance $G_{e-ph}$, determined from the derivative of data in (a). Circles showing the lowest heating power data points, ie, points where $T_{e}$ is very close to $T_{ph}$. Solid line from Ref.~\onlinecite{prunnila:2005}. Dashed line calculated $\partial F_{A}/\partial T_{e}$ and dash-dotted and dotted lines calculated $(\partial F_{S}/\partial T_{e})\times 300$ in 3 and 2 dimensions, respectively. See text for further details.}
\label{fig2}
\end{figure}

The main results of the paper are presented in Fig.~\ref{fig2}. Figure~\ref{fig2}(a) shows the electron temperature $T_{e}$ measured with the thermometer junctions as a function of applied heating power at different bath temperatures. One can readily see that with the same power the strained sample is heated to much higher temperatures than the control. This indicates that the \textit{e-ph} coupling has indeed been reduced in the strained sample as compared to the unstrained control.

To obtain more quantitative results we extract the \textit{e-ph} thermal conductance $G_{e-ph}=\partial P/\partial T_{e}$. This is plotted in Fig.~\ref{fig2}(b) as small dots. The difference between conductances in the control and in the strained sample is about a factor of 20 at the lowest temperatures and a factor of 50 at 400 mK. We have assumed that the phonon temperature $T_{ph}$ equals the bath temperature. As this assumption might break down when the temperature difference between the electrons and phonons becomes large, we plot, as larger circles, the lowest heating power data points ($P\sim0$), where this assumption should be most reliable. We also show as a reference the result from a comparable unstrained sample from Ref.~\onlinecite{prunnila:2005}. Note, however, that in Ref.~\onlinecite{prunnila:2005} the measurements extended only up to 350 mK, the samples had a dopant concentration of $3.5 \times 10^{19}$cm$^{-3}$ and they were fabricated on silicon-on-insulator wafer.

In order to compare these results with the theory, we assume that the electron and phonon sub-systems are each in separate internal equilibrium at temperature $T_{e}$ and $T_{ph}$, respectively. The total energy loss rate (or power loss) $P$, which is equivalent to the heating power, can then be quite generally described by a symmetric formula 
\begin{equation}
P=\mathcal{V}(F(T_{e})-F(T_{ph})),  \label{eq:ELR}
\end{equation}
where $\mathcal{V}$ is the volume of the electron gas. Hence the electron-phonon thermal conductance is given by $G_{e-ph}=\partial F/\partial T_{e}$. According to the mean-field analysis of Ref.~\onlinecite{prunnila:2007}, in semiconductors, the energy loss rate function $F$ can be expressed as a sum of two components 
\begin{equation}
F(T)=\ F_{S}(T)+F_{A}(T).  \label{eq:F_T_semic}
\end{equation}
In single-valley systems only the symmetric part $F_{S}$ exists. Certain lattice perturbations lift the valley degeneracy in many-valley systems and this gives rise to the asymmetric part $F_{A}$. In the diffusive limit, where the electron mean free path ($l_{e}$) is smaller than the inverse thermal phonon wave number ($q_{T}=\frac{k_{B}T}{\hbar v_{\lambda }}$), we can write \cite{prunnila:2007}
\begin{subequations}
\label{eq:F_both}
\begin{eqnarray}
F_{S}(T) &=&K_{2d+1}\frac{8\left( 3/2\right) ^{d}}{9l_{e}\kappa ^{2(d-1)}}%
\sum_{\lambda }v_{\lambda }\left\langle \Xi _{S}^{2}\right\rangle
q_{T}^{2(d+1)}  \label{eq:F_S} \\
F_{A}(T) &=&K_{5}l_{e}\frac{\gamma }{\overline{\gamma }}\sum_{\lambda
}v_{\lambda }\left\langle \Xi _{A}^{2}\right\rangle q_{T}^{6}.
\label{eq:F_A}
\end{eqnarray}
\end{subequations}
Here $K_{m}=\frac{\hbar \nu }{2\pi ^{2}\rho v_{F}}B_{m}$, where $\nu $, $\rho$, and $v_{F}$ are the density of states (for electrons), mass density of the crystal and Fermi velocity, respectively. Numerical factor $B_{m}$ $=\int dxx^{m}/(e^{x}-1)$, the screening wave number (inverse of the screening length) $\kappa =2\left( e^{2}\nu /4\varepsilon_{b}\right)^{1/(d-1)}$ and $\varepsilon _{b}$ is the background dielectric constant. The dimensionality of the electron system is defined through the parameter $d=2,3$. Note, however, that $F_{A}$ does not depend explicitly on $d$. $\gamma $ and $\overline{\gamma }$ denote the total scattering rate and total intervalley scattering rate, respectively ($l_{e}=v_{F}/\gamma $). Effective deformation potential constants $\left\langle \Xi _{S,A}^{2}\right\rangle $ and phonon velocity $v_{\lambda }$ depend on mode index $\lambda $. Here $\lambda =L$ denotes longitudinal and $\lambda =S$ denotes transverse mode. If we assume symmetry of the Si conduction band valleys and bulk phonon system with isotropic acoustic modes we have $\langle \Xi _{S}^{2}\rangle =\widetilde{\Xi }^{2}(\frac{4}{15}\Xi _{u}^{2})$ and $\langle \Xi _{A}^{2}\rangle =\frac{8}{15}\Xi _{u}^{2}(\frac{4}{5}\Xi _{u}^{2})$ for $\lambda =L(S)$, where $\widetilde{\Xi }^{2}=\frac{4}{3}\Xi _{d}\Xi_{u}+2\Xi _{d}^{2}+\frac{2}{5}\Xi _{u}^{2}$. Here $\Xi _{d(u)}$ is the dilatational (uniaxial) deformation potential constant. \cite{herring:1956} We have assumed valley degeneracy of two (six) for $\left\langle \Xi _{S(A)}^{2}\right\rangle$.

Note that in Eqs.~(\ref{eq:F_both}) $F_{S}\propto 1/\kappa ^{2(d-1)}$, whereas $F_{A}$ does not exhibit any dependency on the screening wave number $\kappa $. This reflects the quite general principle according to which $F_{S}$ is strongly screened (at low temperatures) and $F_{A}$\ is unscreened.\cite{prunnila:2007} Lack of screening in $F_{A}$ follows physically from the many-valley picture, where a phonon can create different perturbation to different valleys. Therefore, at low temperatures we should always have $F_{A}\gg F_{S}$ in a many-valley system and hence $F\simeq F_{A}$. If valleys with different symmetry can be depopulated then $F_{A}=0$ and we should observe a large reduction in $F$, because now $F=F_{S}$.

Consequently we would expect the control sample to have $G_{e-ph}\simeq \partial F_{A}/\partial T_{e}$, whereas in the strained sample we should have $G_{e-ph}=\partial F_{S}/\partial T_{e}$. Figure~\ref{fig2}(b) shows these theory predictions as lines. All fixed parameters of the electron system(s) are obtained from standard single electron formulas. For the other parameters we use typical values for Si: $\rho =2.33\times 10^{3}$ kgm$^{-3},v_{T(L)}=4700$ $(9200)$ m/s and $\Xi _{d(u)}=-11.7(9.0)$ eV.\cite{Hull1999} $F_{A}$ has one fitting parameter $\gamma /\overline{\gamma}$, to which we get a value $1.27$ from a fit to the control sample data. The fit was performed in the temperature range $200-270$ mK. In this range the data follows $G_{e-ph}\propto T^{5}$ behavior ($F\propto T^{6}$) and fits well to the theory, but above $300$ mK there is some deviation from the expected $T^{5}$ behavior. The origin of this deviation is not clear, but it is worthwhile noting that Eqs. (\ref{eq:F_both}) are asymptotic zero Kelvin expressions for bulk phonon spectrum.

The strained case ($F_{S}$) has no free parameters and we have plotted ~$G_{e-ph}=\partial F_{S}/\partial T_{e}$ with dimensionality $d=2$ and $d=3$. These theoretical predictions fall significantly below our experimental data and in order to fit everything into same axis we have scaled $\partial F_S/\partial T_e$ by a factor of 300. Here $d=2$\ means a quasi 2D-electron system where density response and screening are essentially 2D due to large thermal phonon wave length (that exceeds the electron layer thickness $t$), which we expect to be a better description of the system at hand. In this case the quasi 2D density of states $\nu _{2}$ is approximated by $\nu _{2}=t\nu _{3}$, where $\nu _{3}$ is the 3D density of states. From Fig. \ref{fig2}(b) we observe that the theory is at least two orders of magnitude below the experimental data irrespective of the choice of $d$. In other words, the theory suggests that by introducing the strain we should have a factor of\ several $1000$ decrease in $G_{e-ph}$, \ but we observe only a factor of $20-50$ decrease in the temperature range $T=200-480$ mK. On the other hand, the experimental $G_{e-ph}$ of the strained sample follows the slope of the 2D theoretical curve ($G_{e-ph}\propto T^{5}$) , which suggests that by introducing a factor of \ $\sim 30$ larger deformation potential constant(s) we would be able to explain the behavior of the strained sample. This would mean that the deformation potential constants should depend strongly on the strain, because the control sample fits to the theory so well (below 300 mK). We are not aware of strain affecting the deformation potential constants, but our data suggest the e-ph coupling mechanisms in heavily doped semiconductors requires further investigation.

In conclusion, we have shown that by inducing strain to a degenerately doped Si layer, we have reduced its electron-phonon coupling by a factor of $20-50$ in the temperature range $200-480$ mK. Our results are in qualitative agreement with theory \cite{prunnila:2007}, but quantitatively the effect is smaller than expected. The reduced electron-phonon coupling in the strained Si might be beneficial for bolometric sensors and cooling applications.

We acknowledge J. P. Pekola, M. Meschke and J. Ahopelto for fruitful discussions. The project has been financially supported by EPSRC through grant EP/F040784/1 and by EU through Projects No. 216176 NANOPACK, No. 256959 NANOPOWER and FP7 Nanofunction Network of Excellence (EC Grant 257375).


\begin{thebibliography}{99}

\bibitem{ridley:1991rev}
B.~K.~Ridley, Rep. Prog. Phys. \textbf{54}, 169 (1991).

\bibitem{giazotto:2006rmp}
F. Giazotto, T. T. Heikkil\"{a}, A. Luukanen, A. M. Savin, and J. P. Pekola, Rev. Mod. Phys. \textbf{78}, 217-274 (2006).

\bibitem{ganthmakher:b1}
V. F. Gantmakher and Y. B. Levinson, \textit{Carrier Scattering in Metals and Semiconductors} (North-Holland Physics Publishing, Amsterdam, 1987). 
 
\bibitem{prunnila:2005}
M. Prunnila, P. Kivinen, A. Savin, P. T\"{o}rm\"{a}, and J. Ahopelto, Phys. Rev. Lett. \textbf{95}, 206602 (2005).

\bibitem{prunnila:2007}
M. Prunnila, Phys. Rev. B \textbf{75}, 165322 (2007).

\bibitem{Vogelsang:1993}
T. Vogelsang and K. R. Hofmann, Appl. Phys. Lett. \textbf{63}, 186 (1993).

\bibitem{savin:2001}
A. M. Savin, M. Prunnila, P. P. Kivinen, J. P. Pekola, J. Ahopelto, and A. J. Manninen, Appl. Phys. Lett. \textbf{79}, 1471 (2001).

\bibitem{herring:1956}
C. Herring and E. Vogt, Phys. Rev. \textbf{101}, 944 (1956).

\bibitem{Hull1999}
R. Hull, ed., \textit{Properties of crystalline silicon} (The institution of Electrical Engineers, London, 1999).

\end{thebibliography}
\end{document}